\documentclass[lnicst,sechang,a4paper]{svmultln}
\usepackage{amssymb}
\usepackage{graphicx}

\usepackage{url}
\urldef{\mailsa}\path|{Khaled.Alanezi, Xinyang.Zhou, Lijun.Chen, mishras}@colorado.edu|
\usepackage[pdfpagelabels,hypertexnames=false,breaklinks=true,bookmarksopen=true,bookmarksopenlevel=2]{hyperref}
\usepackage[outdir=./]{epstopdf,epsfig}

\usepackage{mathtools}
\usepackage{listings}
\usepackage{trackchanges}
\usepackage{tabulary}

\newcommand\comment[1]{}

\begin{document}

\mainmatter  % start of an individual contribution

\title{Panorama: A Framework to Support Collaborative Context Monitoring on Co-Located Mobile Devices}

\titlerunning{Collaborative Context Monitoring on Co-Located Mobile Devices}

\author{Khaled Alanezi \inst{1} \and Xinyang Zhou \inst{2} \and Lijun Chen \inst{1,2} \and Shivakant Mishra\inst{1}}

\authorrunning{Khaled Alanezi et al.}

% the affiliations are given next
\institute{Dept. of Computer Science, University of Colorado, Boulder, CO, USA
\and
Interdisciplinary Telecom Program, University of Colorado, Boulder, CO, USA
\mailsa 
}

\toctitle{Lecture Notes in Business Information Processing}
\tocauthor{Alanezi}
\maketitle

\begin{abstract}
A key challenge in wide adoption of sophisticated context-aware applications
is the requirement of continuous sensing and context computing.
This paper
presents Panorama, a middleware that identifies collaboration opportunities to offload context computing tasks to 
nearby mobile devices as well as cloudlets/cloud. At the heart of Panorama 
is a multi-objective optimizer that takes into
account different constraints 
%of available computing opportunities
such as access cost, computation capability,
access latency, energy consumption and data privacy,
%Based on these constraints, Panorama
and efficiently computes a collaboration plan optimized simultaneously for
different objectives such as minimizing cost,
energy and/or execution time.  
Panorama provides support for discovering nearby
devices and cloudlets/cloud, computing an optimal collaboration plan,
distributing computation to 
participating devices, and getting the results back. The paper
%describes the design and implementation of Panorama and
provides an extensive evaluation of Panorama via two representative context
monitoring
applications over a set of Android devices
and a cloudlet/cloud under different constraints.

%maximizing network lifetime, and maximizing computation speedup.
%A prototype of Panorama has been implemented for Android operating system.
\end{abstract}


\section{Introduction}
\label{sec:introduction}

\comment{
context aware apps and need for offloading
cloud cloudlet mobile devices - work has been done 
key issue: how to obtain the right combination for offloading
Number of factors: security, energy, latency, mobility, willingness to participate, etc

This paper provides a middleware framework that provides 
- support for discovering the available devices for collaboration
- computes the right combination taking into a a variety of factors and multiple goals
- support for parallel and sequential tasks

Contributions
optimization algorithm
support for parallel and sequential
implemented and evaluated under a variety of scenarios
}

In the field of context-aware computing, a wealth of clever mobile applications that monitor user environment to detect and react to events of special interest have recently been proposed; see, e.g., 
\cite{wang2014studentlife,you2012carsafe,lu2012stresssense}.
However, a major obstacle towards wide adoption of context-aware applications is the requirement of continuous context monitoring. User context can change at any time and it is crucial for the application to detect those changes promptly. This requirement is difficult to accommodate due to limited smartphone resources, particularly the battery resource. Moreover, despite significant advances in smartphone processing power, context computation latencies remain prohibitively high for several
interesting applications such as cognitive assistance
%, particularly when
%compared with computations in the cloud 
\cite{satyanarayanancloudlets}.
For these reasons, users tend to avoid using context-aware applications.

Different offloading techniques have been proposed recently to
address these issues of limited battery life and long computation latency.
These techniques fall into two broad categories. In the first category, resource-hungry tasks are off-loaded to
powerful servers residing in the cloud, leading to 
both computation speedup and energy efficiency. However, accessing the
cloud incurs additional cost for the user in terms of cloud access fee
and cellular data plan. In addition, access latency for cloud can be quite
high.
%Another problem is the high communication latency for accessing the cloud, which
%hinders usability of the applications.
%For example, in a recent vision paper \cite{miluzzo2012vision}, researchers from AT\&T have raised a concern about the ability of the wireless Internet providers to handle increased traffic generated from offloading speech, image and audio tasks to the cloud. 
%, specifically cognitive assistance applications. 
To address this, researchers are introducing cloudlets,
acting as a middle-tier to bridge the gap between the mobile devices and the cloud
\cite{satyanarayanancloudlets}.

In the second category, mobile applications use nearby mobile devices to share tasks, thereby minimizing the need for accessing cloud resources \cite{conti2010opportunities,lee2012comon}. %We see that this form of collaborative computing among nearby mobile devices can result in several important benefits for context-aware applications.
This helps with avoiding cloud and ISP charges, as  nearby resources can be
personal devices, or mobile devices of family members and coworkers.
% who are willing to share their resources, thus avoiding a need to access costly cloud resources.
This also eliminates redundant sensing and computation, if several nearby
mobile devices are interested in the same (shareable)
context~\cite{lee2012comon}.
%, they can collaborate to share different parts of the context monitoring task. A variety of spatial and social contexts have been identified as sharable \cite{lee2012comon}. Examples include location, temperature, noise level, conversation, lecture, dining, etc. 
In addition, collaborative context monitoring extends sensor modalities and tackles the smartphone position problem~\cite{khaled2013mobiquitous}.
However, this technique suffers from
uncertainties due to the ad hoc nature of the network, lack of 
any apparent incentives for participation, security and privacy, varying device
capabilities and device mobility.
\comment{
Some mobile devices may be incapable of generating the required context due to the absence of a required sensor (e.g., CO\textsubscript{2} sensor), while some devices may be in unfavorable position \cite{khaled2013mobiquitous} rendering them useless for sensing.
%, e.g. a mobile device inside a bag can't take a picture of the surrounding environment. 
With collaboration being enabled, sensing and computation subtasks can be assigned to devices based on their resource availability and current position. 
%Finally, collaborative context sensing can lead to improved performance as well as conserving energy.
}

\begin{figure*}[!t] 
\begin{minipage}[t]{0.5\textwidth}
\raisebox{0.2\height}{\includegraphics[width=\textwidth]{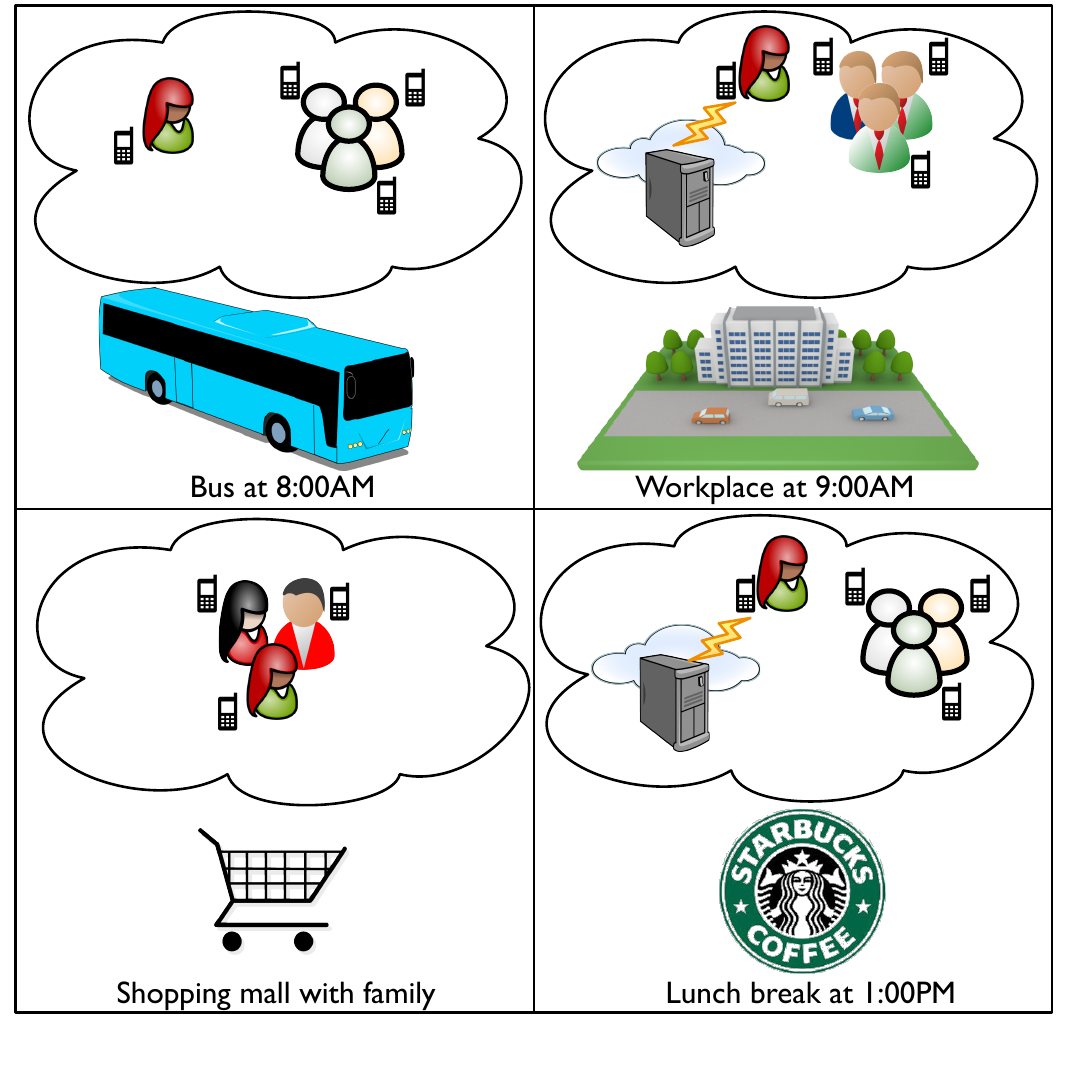}}
\caption{Alice's Typical Day}
\label{fig:typical} 
\end{minipage} \hfill
\begin{minipage}[t]{0.47\textwidth}
\includegraphics[width=\textwidth]{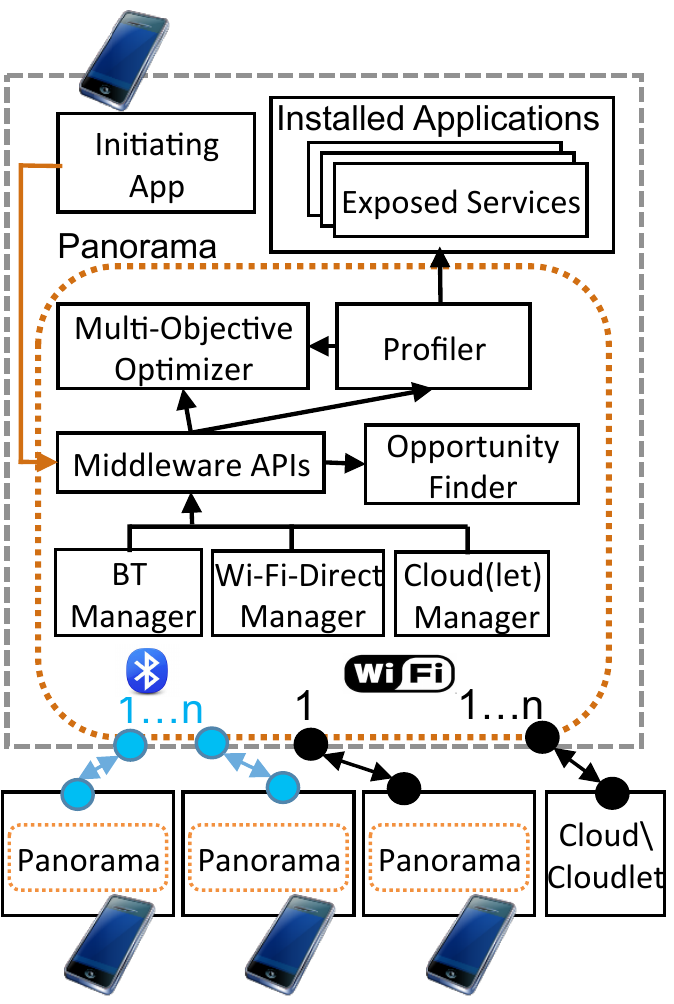} 
\caption{Panorama's Architecture}
\label{fig:architecture} 
\end{minipage} \hfill
\end{figure*}

It is clear that both of these offloading techniques have their pros and
cons with one of them suitable for one scenario and the other one for a
different scenario.
At present, offloading solutions to cloud, cloudlet or
nearby mobile devices exist in isolation.
%For example, a solution that considers offloadling to cloud does not take
%advantage of utilizing nearby mobile devices or cloudlets to further reduce
%cost, execution time, and/or energy consumption. 
With proliferation of mobile
devices, increased availability of (nearby) computing servers that can operate
as cloudlets, and improved connectivity to the cloud, a highly likely scenario
is one where a user has access to multiple computing resources whenever she/he
needs to perform a context computation. Figure~\ref{fig:typical} illustrates
four common scenarios in a typical user's (Alice) life. In the morning,
Alice takes a bus to go to her work ({\it Bus scenario}). During her bus ride,
she can perform collaborative computing with mobile devices of other
bus riders. Later, in her work place ({\it Work Place scenario}), she can
perform collaborative computing with mobile devices of co-workers as well as an
office server (cloudlet) accessible within one network hop. During lunch time
({\it Lunch Break scenario}),
Alice goes to a restaurant, where she can perform collaborative computing with
mobile devices of other restaurant customers as well as a cloudlet provided 
by the restaurant. Finally, in the evening or on weekends, Alice goes for 
shopping in a mall with her family members and friends
({\it Shopping Mall scenario}), where she can use their mobile device
for collaborative
computing.
%At present, the user however is
%limited to
%using either only the nearby mobile devicess, or the cloudlet, or the cloud.

In this paper, we present a middleware framework called {\em Panorama} that enables
mobile applications to reap the benefits of every computing opportunity
(cloud, cloudlets, and other mobile devices) available at runtime.
Panorama runs on multiple mobile nodes,  
and builds an optimized collaboration plan
taking into consideration the users' performance objectives and the participants' preferences and constraints.
A key challenge addressed in Panorama is to ensure an optimal partitioning of
the computation task among available mobile devices, cloudlet, and cloud. 
Panorama considers important and practical constraints in collaboration
planning, such as 
%the mobility of mobile devices,
the energy constraints of mobile devices, and their computation and
communication capabilities. It also 
considers the costs involved in accessing
% cloudlets/cloud and using neighboring mobile devices, as well as 
%the latencies in accessing 
nearby mobile devices vs. cloudlets/cloud. It also takes into consideration
security, privacy, and
trust relationship of the participating devices. 
At the heart of collaboration planning in Panorama is a versatile
multi-objective
optimization framework that takes into account various constraints of available computing opportunities and efficiently computes a collaboration
plan that optimally trades off different performance objectives such as
minimizing the overall cost, minimizing the energy consumption, and minimizing the execution time. Panorama provides support for both parallel and  sequential (pipeline) task structures, two most common structures in context monitoring applications. 

We have prototyped and extensively evaluated Panorama under a variety of
scenarios in the presence of several different
network configurations of mobile devices, cloudlet and cloud and under different device constraints. 
We have experimented with two representative context-aware applications: speech recognition (a parallel task) and ambiance sound monitoring (a sequential task). Experimental results show that Panorama can achieve both reduced computation time and decreased energy consumption while working within the constraints set by the collaborators, such as limits on the contributed energy, cost budget, and privacy requirement. Experimental results also show that Panorama is expressive and flexible in realizing different tradeoffs between completion time, energy consumption, and/or cost. 
%Our prototyping and evaluation have demonstrated definitely the effectiveness and efficacy of Panorama in enabling optimal collaboration for context-aware applications.
Panorama is completely automated with no user intervention needed after installation. A device with Panorama can automatically join a collaboration network when needed, and run tasks that are suitable for it.

%Third, Panorama addresses a very important problem in collaborative computing, namely how to learn about collaboration opportunities without consuming too much power. To this end, Panorama utilizes Bluetooth names of potential collaborators as a means of communicating information required to perform the required optimizations with minimal engagement from potential collaborators side. 

%We have prototyped and extensively evaluated Panorama under a variety of scenarios in the presence of several different
%network configurations of mobile devices and cloudlet and under different device constraints. Panorama is completely automated, and no user intervention
%is required after it is installed and started. The device with Panorama
%running can automatically join a collaboration network when needed, and
%execute tasks that are suitable for it. Our experimental results have demonstrated definitely the effectiveness and efficacy of Panorama in enabling 

\section{Design}
\label{sec:design}

\subsection{Overall Architecture}
\label{sec:arch}
Panorama is designed according to the current mobile application development standards, and can be easily adopted without incurring much change to the current mobile software stack. 
%Panorama is a middleware that runs on mobile nodes and acts as a bridge between them. 
It provides APIs to allow applications to discover nearby devices, cloud and cloudlets, build a network, and delegate tasks to them. It also provides APIs to allow other mobile devices to discover local resources and to accept task delegation. The overall architecture of Panorama is shown in Figure~\ref{fig:architecture}. A device acts as an initiator and triggers the network creation phase when it needs to compute a costly context and is looking for collaborators. Panorama's design supports diverse network interfaces like Bluetooth, Wi-Fi Direct, and connections with IPs where cloudlets/cloud reside. Bluetooth standard allows creation of Piconets where the initiating device connects to multiple devices in a star topology. The initiating device can connect to another device using WiFi-Direct and to previously defined IPs for cloudlets/cloud.

Panorama is implemented as a background service that exposes the middleware APIs to applications looking for collaboration opportunities.
The main component is the {\it middleware APIs component}. This component contains the APIs that the applications can call to use the framework's services.
The {\it Bluetooth Manager} and {\it Wi-Fi Direct Manager} components implement technical details of short-range communication channels. Panorama defines the behavior that every communication channel must provide to support task collaborations like searching for other devices, connecting to other devices, accepting connections from other devices, accepting and responding to resource inquiry messages, and finally accepting and processing task delegation. Panorama currently supports two communication interfaces with other mobile nodes: Bluetooth and Wi-Fi Direct. With this design, it would be easy to plug in new communication interfaces (e.g., NFC or ZigBee) without any significant change in Panorama.

The {\it Cloud(let) Manager} component implements technical details of connecting with cloudlets/cloud. Currently, we pre-configure the IP addresses where the implementation for specific context monitoring task exists. However, we expect that network resource discovery techniques can be utilized to discover cloudlets/cloud efficiently.

The {\it Profiler} component gathers collaboration-relevant information about the mobile node Panorama is running on and provides it to other nearby mobile devices through Panorama's APIs. 
%It gathers information that are important for making task collaboration decisions. As we will see later, 
The initiating node delegates different context sub-tasks to different devices based on the profiler information. Currently, this component provides two types of information: available set of services on the device along with their
performance metrics and constraints of the mobile device. Available services
here refer to context derivation code that applications expose so that other mobile devices can execute their context-aware task on the device. In the current
design, the required code should be available on the device before collaboration can take place. Device constraints include energy quota, time quota, and incentives, etc. %\khaled{reference section -partitioning- in implementation for more details}

The {\it Multi-Objective Optimizer} component 
employs multi-objective optimization to find the best collaboration plan that conforms to the device constraints and achieves the initiator's objectives. Currently, we optimize for energy consumption, execution time, and cost. However, due to its flexibility, the optimization model can be easily extended to accommodate for other parameters when required. 
Details of this component are discussed in Section \ref{sec:optimization}.
Finally, the {\it Opportunity Finder} component performs regular scanning
for nearby devices using Bluetooth and maintains a list of recently discovered devices to be utilized whenever collaboration is required.

\subsection{Application Partitioning and Profiling}
\label{sec:partition}

Collaborative context monitoring involves changing the application execution model from standalone execution on a single mobile device to distributed execution on multiple mobile devices, clouds and cloudlets. This requires partitioning the application and making the code for calculating context available on collaborating nodes before the collaboration. Panorama's design requires the context code to be available on other mobile devices before the collaboration. This design choice is reasonable since Panorama targets shareable contexts that will be of common interest. For example, a programmer will write a speech recognition component that takes an audio as input and returns text as a result. This component can then be made available for other applications running on the device as well as for nearby collaborators by being exposed as a service through the operating system. Note that this design choice is consistent with research in the field that proposes contextual data units \cite{chu2011mobile} and envisions sharing them among collaborators \cite{de2015group}. Technically, Panorama utilizes the Android service component to support this design (see Section \ref{sec:implementation}). For servers, we envision a future where popular shareable context is exposed by server APIs analogous to web APIs.

For application profiling, Panorama tracks the required execution time and energy consumption for exposed services and provides this information through the API to other mobile devices. Panorama uses information gathered from previous invocations to build a linear regression model similar to the work in \cite{cuervo2010maui} that predicts the execution time and energy consumption for future tasks delegated to the node. We choose to use file size as the input to this regression model. This choice has proven accurate for speech recognition tasks in our current implementation. The energy profiling is done manually by taking measurements from an external power source. This workaround solution is required due to the lack of an accurate API that exposes energy consumption of the device to solutions like Panorama.

\subsection{Optimization Models}
\label{sec:optimization}

%In this subsection, we describe the optimization models implemented in Panorama's Multi-Objective Optimizer component. These models are used to make task partitioning and collaboration plan optimized for different performance objectives under various participant preferences and resource restrictions. 
 
Consider a system where $n$ devices, indexed by $i=1,\cdots,n$, collaborate on a certain task with a total workload of $w$. Let $x=(x_1,\cdots, x_n)$ denote the allocation of the task, with device $i$ being allocated an amount $x_i\geq 0$ of workload. Obviously, $\sum_{i=1}^n x_i=w$. Denote by $e_i$ the energy consumption for processing one unit of workload by device $i$. We assume that each device $i$ has an energy budget $b_i$ that it is willing to expend for collaboration, i.e.,  $e_i x_i \leq b_i$. Denote by $c_i$ the payment received by device $i$ for processing a unit of workload, and $B$ the initiator's total budget on payment. So,  $\sum_{i=1}^n c_i x_i\leq B$. We further assume that each device $i$ takes an amount $f_i$ of time to process one unit of workload. We aim to minimize both the energy consumption and the completion time, which is formulated as the following multi-objective optimization problem:
{\small
\begin{eqnarray}
\min_{x \succeq 0}~(\mbox{w.r.t}~R^n_+)&&  (\sum_{i=1}^n e_i x_i,~~\max_i\{f_i x_i\})\label{eqn:obj1}\\
\mbox{s.t.} %&& x_i\geq 0, i=1,\cdots,n\\
&&e_i x_i \leq b_i, i=1,\cdots,n\\
&&\sum_{i=1}^n x_i=w\\
&&\sum_{i=1}^n c_i x_i\leq B. \label{eq:coneb}
\end{eqnarray}
}

%\hspace{-1mm}
Introducing a weight $\gamma\geq 0$ to specify the tradeoff between energy consumption and execution time, we can solve the above problem by scalarization, which can be reformulated as a linear program (LP):
{\small
\begin{eqnarray}
\min_{x\succeq 0} &&  \sum_{i=1}^n e_i x_i + \gamma t \label{eqn:obj2}\\
\mbox{s.t.} %&&x_i\geq 0, i=1,\cdots,n\\
&&e_i x_i \leq b_i, i=1,\cdots,n\\
&&f_i x_i \leq t, i=1,\cdots,n\\
&&\sum_{i=1}^n x_i=w,~~~\sum_{i=1}^n c_i x_i\leq B. \label{eq:coneb1}
\end{eqnarray}
}
\hspace{-1.5mm}A larger (smaller) $\gamma$ means a higher preference/priority on short completion time (low energy consumption). %and a smaller $\gamma$ means a higher preference/priority on low energy consumption. 
In practice, the value of $\gamma$ is set based on the initiator's preference. 

Notice that in the above optimization problems, we impose a hard constraint on the initiator's budget; see equation \eqref{eq:coneb}. But we can also make the payment an objective to optimize. For example, we can optimize both the initiator payment and the completion time under the energy budget constraint, i.e.,  
\comment{
{\small
\begin{eqnarray}
\min_~{x\succeq 0} (\mbox{w.r.t}~R^2_+)&&  (\sum_{i=1}^n c_i x_i,~~\max_i\{f_i x_i\})\label{eqn:obj3}\\
\mbox{s.t.}&&x_i\geq 0, i=1,\cdots,n\\
&&e_i x_i \leq b_i, i=1,\cdots,n\\
&&\sum_{i=1}^n x_i=w. \label{eq:conwl}
\end{eqnarray}
}
\hspace{-1mm}This again can be solved by scalarization:
}
{\small
\begin{eqnarray}
\min_{x\succeq 0} &&  \sum_{i=1}^n c_i x_i +\gamma t \label{eqn:obj4}\\
\mbox{s.t.}%&&x_i\geq 0, i=1,\cdots,n\\
&&e_i x_i \leq b_i, i=1,\cdots,n\\
&&f_i x_i \leq t, i=1,\cdots,n\\
&&\sum_{i=1}^n x_i=w. \label{eq:conwl1}
\end{eqnarray}
}

The above modeling framework can be easily extended to incorporate different performance objectives or concerns. For example, for certain reason such as privacy concern, we may require certain portion of workload $u$ to be processed at a subset $i=1,\cdots, m$ of devices such as those that can be trusted. 
%This can be ensured by imposing an additional constraint $\sum_{i=1}^m x_i = u$, which leads to the following optimization problem if the initiator aims to optimize for the energy consumption and completion time:
This can be ensured by imposing an additional constraint $\sum_{i=1}^m x_i=u$ to the above optimization problems. 
%{\small
%\begin{eqnarray}
%%\min_x &&  \sum_{i=1}^n e_i x_i + \gamma t \label{eqn:obj5}\\
%%\mbox{s.t.}&&x_i\geq 0, i=1,\cdots,n\\
%%&&e_i x_i \leq b_i, i=1,\cdots,n\\
%%&&f_i x_i \leq t, i=1,\cdots,n\\
%&&\sum_{i=1}^m x_i=u. %~~~\sum_{i=1}^n x_i=w. \label{eq:conp}
%%&&\sum_{i=1}^n c_i x_i\leq B. \label{eq:coneb1}
%\end{eqnarray}
%}
%The above problem has been implemented in our experiment incorporating privacy concern in Section ``Evaluation''. 

%\hspace{-1mm}
In practice,  the values of $e_i$ and $f_i$ can be measured/estimated as described in Section \ref{sec:partition}. The value of $c_i$ will be determined by each device/collaborator based on its resource scarcity or abundance as well as incentive. The total number $n$ of collaborating devices is usually a small number less than 10, resulting in a small LP problem. The LPs \eqref{eqn:obj2}-\eqref{eq:coneb1} and \eqref{eqn:obj4}-\eqref{eq:conwl1} can be solved on smartphone using existing LP solvers, e.g., Apache for Java, in tens of millisecond. We have implemented a customized solver especially for Panorama. 
%However, since it is rare to have a large number of collaborations around, and also it's unsafe and inefficient to divide and spread a task to too many devices, we don't consider large $n$ here.

\subsection{Discovery Protocol}
\label{sec:discovery}
For collaboration, Panorama needs to know what devices are available nearby and how long those devices are expected to stay within the collaboration range. Under Bluetooth v3.0, Panorama needs to scan its surroundings regularly, which results in significant energy overhead. To minimize this overhead, Panorama utilizes adaptive scanning based on discovered number of peers as described in \cite{han2012ediscovery}. The new Bluetooth Low Energy (BLE) protocol provides lightweight mechanism for broadcasting device capability beacons in a connectionless mode called advertising. Panorama can leverage this feature for efficient service discovery and switch to classic Bluetooth for sending files at higher rates. Unfortunately, none of our Android devices support BLE peripheral mode required for advertising. We plan to incorporate BLE in Panorama as part of future work. %\khaled{suggestion: add small BLE experiment and say that Panorama can work on two modes? this can also help in explaining how BLE can help in issues like mobility and privacy}

\section{Implementation}
\label{sec:implementation}
We have implemented Panorama as a background service on the Android platform, which can be installed as a user-space application. After installing Panorama, context-aware applications running on the same mobile device can use Android IPC to call its APIs. Panorama's simple interface has a start/stop button that can be used by users to indicate their willingness to engage in collaboration. Once Panorama is started, it can automatically accept Bluetooth connections from co-located devices that also run Panorama. Bluetooth connection between Panorama copies running on different mobile devices can take place automatically without user intervention. In order to support this requirement, Panorama uses a specific Bluetooth UUID as an identifier and connects using Bluetooth insecure channel. This design allows for automatic creation of connections with co-located devices which is a mandatory requirement for systems such as Panorama to work. However, it introduces security risk from an adversary with access to the UUID. Techniques to secure mobile ad hoc networks such as reputation systems \cite{buchegger2003robust} and secure key management  \cite{capkun2003self} can be employed to secure Panorama. We are considering implementing these techniques in Panorama as part of future work.
Panorama also utilizes Wifi-Direct as an additional communication channel. However, the connection has to be accepted by the receiving party since this is the only supported scheme on Android implementation of Wifi-Direct. When the communication network is established, devices can discover resources and delegate tasks to each other.

\subsection{Panorama's Programming Interface}
Method signatures of Panorama's APIs are defined using the Android Interface Definition Language.
%\cite{aidl}.
In order for third-party applications developers to call Panorama's APIs, they will need to include a copy from the .aidl file in their application package and bind to Panorama's middleware service. Table \ref{table:api} lists method signatures from this file.
The group of APIs handling Bluetooth, Wifi-Direct and Cloudlet perform the required functions for creating the network using the communication channels. To create the Bluetooth network, a device checks the freshness of recently discovered devices list using {\it get\_bt\_devices}, then invokes the {\it create\_piconet} API. The latter API automatically connects to nearby devices running Panorama.
An application can also connect to a device through WiFi-Direct and a cloudlet/cloud server through Wi-Fi using the Wifi-Direct APIs and cloudlet APIs respectively. For generic APIs, the {\it discover\_nw} API sends a discovery message for all connected devices with the required service name.  Panorama's design utilizes the Android service component for code discovery. That is, application developers will write context derivation code in an Android service and expose it under a unique identifier through the Android OS. Accordingly, whenever a device receives a discovery message, it triggers the {\it get\_local} API, which checks whether the required service is installed on the device by checking exposed services against the provided {\it service\_name} string.
After gathering information about connected devices, the application can trigger the optimization process and perform the collaboration using the {\it perform\_optimization} and the {\it execute\_optimization} APIs respectively. Upon receiving task delegations, collaborating devices can process their portions of the task using the {\it process\_local} API.
\begin{table}
\footnotesize
\caption{Method signatures from Panorama's aidl file}
\begin{center}
\begin{tabular}{ |c|c| } 
  \hline
 Generic APIs & Bluetooth, Wifi-Direct \& Cloudlet APIs \\
  \hline 
 \texttt{List<Device> discover\_nw(service\_name)} & \texttt{List<Device> get\_bt\_devices} \\ 
 \texttt{Device get\_local(service\_name)} & \texttt{create\_piconet} \\ 
 \texttt{Plan perform\_optimization(task)} &  \texttt{start\_wd\_discovery} \\ 
 \texttt{Result execute\_optimization} & \texttt{List<Device> get\_wd\_devices} \\ 
 \texttt{Result process\_local(task)} & \texttt{ connect\_wd(device\_name)} \\
  & \texttt{connect\_to\_cloudlet} \\ 
 \hline
\end{tabular}
\end{center}
\label{table:api}
\end{table} 

%Essentially, the design goal of Panorama is to abstract the details of collaboration planning and execution thereby allowing programmers to focus on their application logic. Nonetheless, exposed APIs can be expanded or minimized according to the desired degree of control over how collaborations can occur. 
In the description above, we have provided detailed APIs from Panorama for clarity. However, we note that these APIs can be combined to shield collaboration logic from the application logic and make task delegations happen automatically. For example, we have implemented an API that both connects to nearby devices and discovers them for a specific service in one step.

\subsection{Experiment Testbed} 
\label{sec:testbed}
To evaluate the utility of Panorama, we have implemented two context-aware applications representing two different application structures: parallel structure and pipeline structure. 
% as shown in Figure \ref{fig:tasks_struct}. 
A speech recognition application that is based on PocketSphinx \cite{huggins2006pocketsphinx} to perform speech recognition from a dictionary %that requires processing of multiple speech files 
represents a parallel task. %We used PocketSphinx \cite{huggins2006pocketsphinx} to perform speech recognition from a general dictionary.
This task is computation-intensive, making it a good candidate for collaboration. For the pipeline structure, we implement the sound ambiance monitoring task from \cite{lu2009soundsense}. We define three stages and run them in three different Android services components to distribute the application. First, an audio recording stage, which represents the sensing stage. Second, a stage that calculates FFT for the audio window and generates features to classify sound as either music or speech. Finally, a third stage that takes the FFT as input and generates MFCC vector, which is then used to identify the gender of the speaker. This is only used when the sound is detected as speech. We also implement both applications using Java to be able to run them on cloudlets and clouds.
We integrate these applications with Panorama and run experiments on four Android mobile devices running different versions of the Android OS and a laptop to emulate a cloudlet compute box. We also rent an Amazon EC2 server to use for experiments involving the cloud. The Android devices used are Galaxy S4, Galaxy Note, Galaxy Tab 3, and Galaxy Nexus. Galaxy Note, Galaxy Tab3  and Galaxy Nexus have dual-core processor while Galaxy S4 has a quad-core processor.

\section{Evaluation}
\label{sec:evaluation}
\subsection{Methodology}
We have evaluated Panorama for a variety of scenarios under different collaboration opportunities, resource restrictions, and incentives. The experimental settings are chosen to reflect real-life scenarios that a system like Panorama may face. Collaboration opportunities include cloudlets/cloud as well as multiple
mobile devices belonging to the user,
his/her friends and family members, and/or strangers. The initiator may have different objectives, and the collaborators may have different constraints in terms of energy, time, cost, and privacy. 
%We will test the utility of Panorama in different scenarios and evaluate how it adapts to different available opportunities, resources restrictions, and participant preferences. 
%For available opportunities, we employed scenarios where a user faces multiple mobile nodes for others, when a user faces multiple mobile nodes for others in addition to a mobile node belonging to her and when the user faces multiple mobile nodes and a cloudlet. Also,

% we implemented several restrictions that the participants might impose on the collaborations and tested the behavior of Panorama under these restrictions. Panorama's generality makes it impossible to cover all the cases that the framework can face. However, the results we show proves its adaptability and utility to various scenarios, thereby, proving that the same behavior will occur with future scenarios. 
 
The execution time reported in the experiments is the total time to execute the required task using a collaboration, including the time for connecting with other devices or cloudlet, devising an optimal task partitioning plan, shipping subtasks, and gathering the results back. The energy consumption reported is the sum of energy consumed in all mobile devices (not cloudlet) that participate in the collaboration. It takes into consideration the energy consumed in all stages from the creation of network for collaboration to the gathering of the results back to the initiating device. 
To measure energy consumption of different activities, we log the device electric current drain indicated by the power supply unit, and then subtract the average current drain observed before the measured activity starts in order to obtain the additional current drain caused by the collaboration activity. We multiply the additional current with the voltage applied at the battery terminals of the mobile device to get the instantaneous power consumption of the activity in Watt and then integrate it over time to obtain the energy consumption of the activity in Joule. We ensure that there are no other applications running in the background.
% during the experiment. 
%So the current measurement before the collaboration is mostly attributed to powering the screen of the mobile device.
We repeated each experiment five times and report the 
average of the measurements from these five trials. We also report 
standard deviation, which is rather low in all experiments.

\subsection{The Utility of Multi-Objective Optimizer}

We use speech recognition to evaluate the adaptability of Panorama to different participant preferences and resource restrictions.
% that is enabled by its Multi-Objective Optimizer component. 
Speech recognition is a good candidate for task collaboration because of its compute-intensive nature, and 
is used as an example context-aware task that can be distributed in parallel. We envision Panorama integrating with context-aware applications that require general speech recognition. %to perform different behaviors. 
In the experiments, we consider a scenario where an audio file of 4 MB is recorded and requires performing speech recognition.

\subsubsection{Tradeoff between Energy Consumption and Execution Time} %Multi-Objective Optimization}
\begin{figure*}[!t] 
\begin{minipage}[t]{0.48\textwidth}
\includegraphics[width=\textwidth]{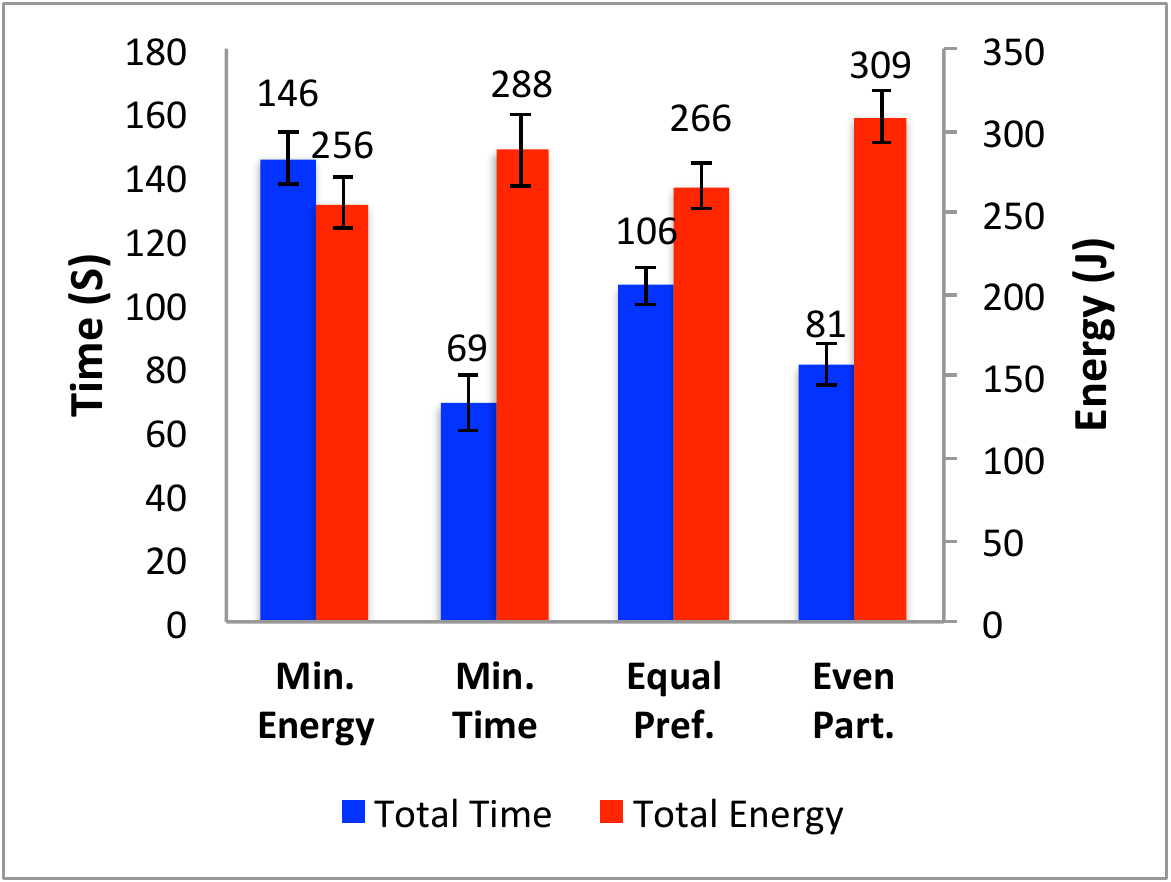} 
\caption{Tradeoff between energy consumption and execution time by Multi-Objective Optimizer (3 collaborators).}
\label{fig:strategy} 
\end{minipage} \hfill
\begin{minipage}[t]{0.48\textwidth}
\includegraphics[width=\textwidth]{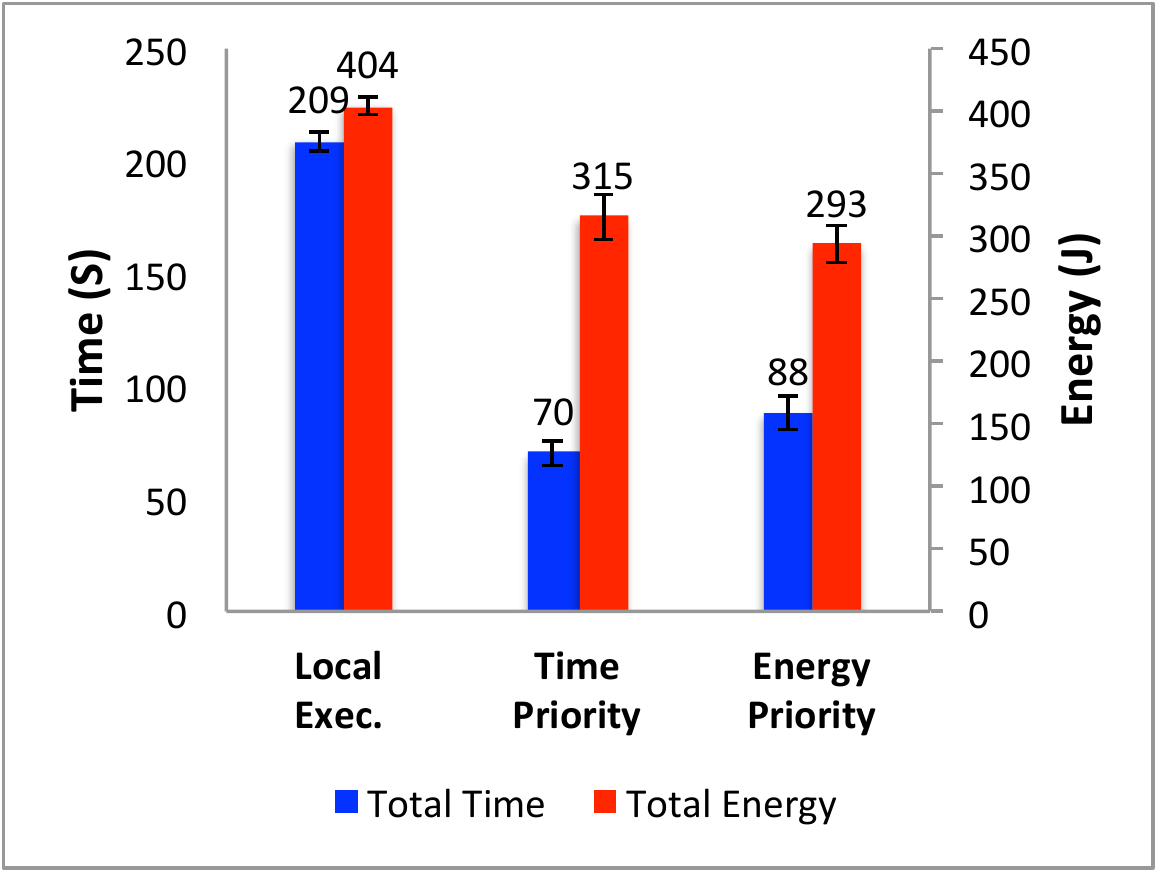} 
\caption{Impact of privacy on energy and time optimization (3 collaborators).}
\label{fig:strategy-privacy} 
\end{minipage} \hfill
\end{figure*}

We first demonstrate that Panorama provides support for appropriate
partitioning to achieve the desired tradeoff between energy consumption and execution
time. Such a tradeoff is needed in a Bus or a Shopping Mall scenario
described in Figure~\ref{fig:typical}, where only mobile devices may
be available for collaboration and the user does not have access to a power
source. Due to the lack of access to power source,
the remaining battery level dictates how important it is to minimize
energy consumption during computation.
%For example, it is reasonable to reduce the overall energy
%consumption at the expense of increased execution time if the remaining
%battery level is low.

In this experiment, we assume that there are two mobile devices
available in the user's vicinity in addition to the user's own mobile device.
We also assume that the speech data does
not contain any sensitive information and so privacy is not a factor in
optimization. In the next experiment, we will take into consideration the privacy concern
when certain parts of the speech data are sensitive in nature. 
Recall from Subsection \ref{sec:optimization}
%\ref{sec:optimization}
that different choices of weight $\gamma$ correspond to different tradeoffs between energy consumption and execution time. 
Figure \ref{fig:strategy} shows the comparison between $\gamma=0$ (minimize energy consumption, corresponding to a situation with relatively low remaining battery level), $\gamma=\infty$ (minimize execution time, corresponding to a situation with relatively high remaining battery level), and $\gamma=1$ (equal preference over energy and time, corresponding to a situation with moderate remaining battery level). We also contrast this with a naive partitioning strategy that divides the task evenly over all the participants. 
Figure \ref{fig:strategy-partition} shows the  corresponding partitioning of speech data for each of the collaborators for these different cases. For brevity, we report in the same figure the file partitioning of speech recognition tasks for other experiments as well that are described later.

We see that in the case of minimum energy (low remaining battery level),
bigger chunks of file are sent to the participating device with low energy consumption without any emphasis on exploiting parallelism to achieve computation speedup, leading to slow execution. The opposite trend is observed for the case of minimum time (high remaining battery level). Here, the total execution time is minimized at the expense of increased energy consumption.
The equal preference case represents a compromise between the previous two cases, with execution time and energy consumption in between those of these two cases. Finally, %due to equal file chunks assignment, 
the even partitioning scenario is able to exploit parallelism to achieve a good performance in time. However, it consumes the most energy because of the lack of any planning in this aspect.
%The equal preference case leads to a good tradeoff between energy and time: compared with the minimum energy case, Panorama uses only $18\%$ more energy but saves $37\%$ time; and compared with minimum time case, Panorama takes $15\%$ more time but saves $14\%$ energy. Compared with the even partitioning case, the equal preference case is better off in both energy and time.  

Notice that in the previous experiment, the task distribution is same whether
the user is in a Bus scenario or a Shopping Mall scenario, because the speech
data does not contain any sensitive information.
We now consider the case where some parts of the speech data contain
sensitive information (1.5 MB out of the total 4 MB is sensitive). In the
Bus scenario, since the mobile devices other than the initiator's are 
untrusted, the sensitive parts of the data cannot be shipped to them. 
As a result, only 2.5 MB of (non-sensitive) speech data is available for
collaboration in this case and the remaining 1.5 MB of (sensitive) data must
be processed at the initiator's device. Figure~\ref{fig:strategy-privacy}
shows the results of this case. We see that in both time and energy priority cases, the achieved time gain from exploiting other mobile nodes is more than $50\%$. In the case of energy priority, $27\%$ energy is saved compared to local execution due to shifting of the (non-sensitive) portion of the task to a more efficient mobile node. Also, when we compare the energy priority and time priority cases, we see that Panorama is able to devise the best plan for executing the non-sensitive part of the file. For the time priority, Panorama divides the file efficiently (see Figure \ref{fig:strategy-partition}) to save $20\%$ of time when comparing to the energy priority case. As for the energy priority, Panorama sends the file chunks to the more energy efficient device thereby saving $7\%$ more energy.
%An interesting observation that we would like to highlight in this experiment is that the time for both energy priority and time priority cases is almost equal. This is due to the fact that the local execution time of the sensitive 2 MB part of the file dominates the execution time. That said, Panorama devised the required plan for faster execution in the case of time priority but the improvement was overshadowed by the time taken by the initiator to compute the sensitive 2 MB locally. %On the other hand, Panorama was able to save energy since a larger part of task was executed by a more energy-efficient device. 
For the Shopping Mall scenario, since all devices are trusted, the
presence of any sensitive data does not make any difference in 
task distribution. The results are same as those reported in
Figure~\ref{fig:strategy}.
%The third case we looked at was distributing the file chunks equally between participants. For this case, we deliberately divided the file into three 2.7MB chunks and distributed them to the three participating node. We note here that this situation produced worse energy consumption even when compared to the best time case. Also, the execution time was bad with 82 seconds more than the best case time. In the last case, we used Panorama's decision making strategy with equal weights assigned to energy and time objectives. We note that Panorama was able to find a compromise solution when compared to best time and energy cases. Compared to the best energy scenario, Panorama's uses only 15\% more energy with the benefit of saving 37\% time savings. As to comparing with best time scenario, Panorama uses 22\% more time bust saves 4\% energy.
%\textcolor{blue}{We need to discuss the details of this experiment}
\subsubsection{Impact of Collaborator Constraints}
\begin{figure*}[!t] 
\begin{minipage}[t]{0.48\textwidth}
\includegraphics[width=\textwidth]{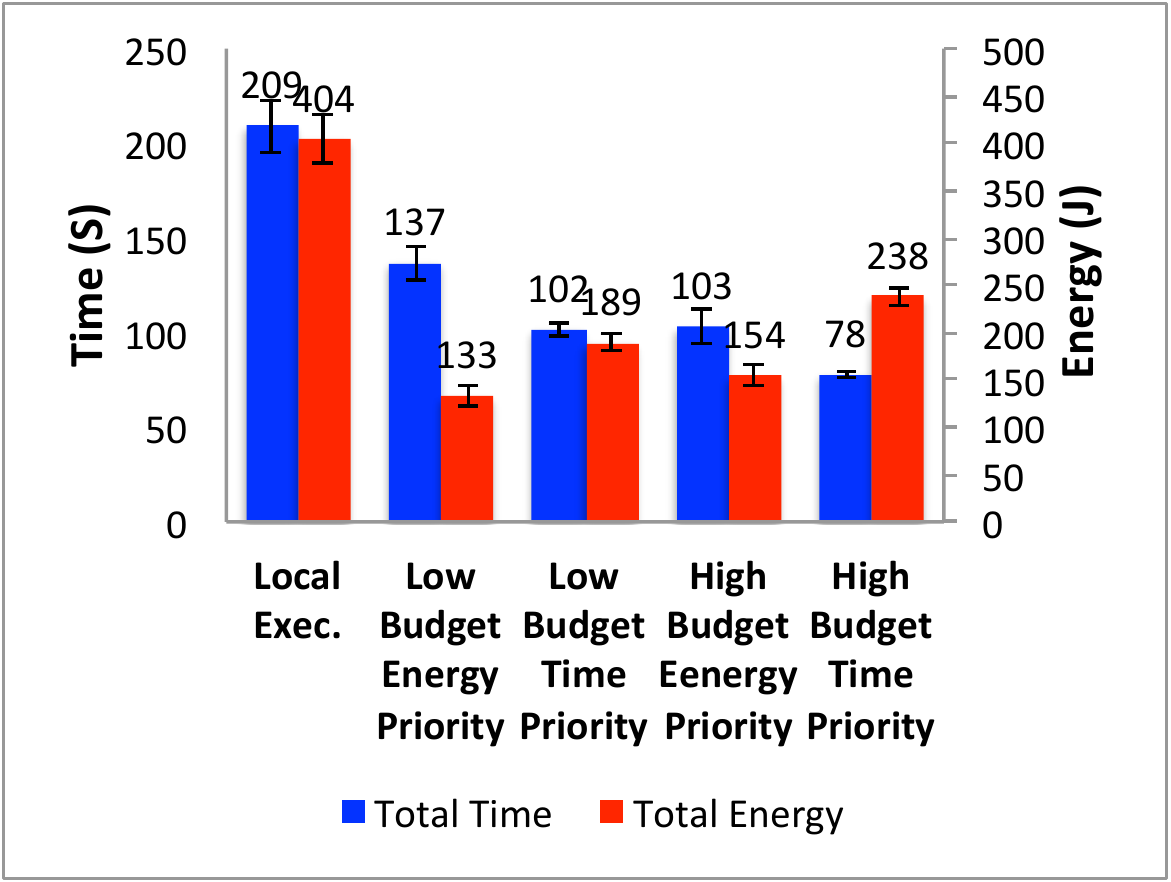} 
\caption{Impact of the initiator's willingness to pay (3 collaborators \& 1 personal low end device).}
\label{fig:incetives} 
\end{minipage} \hfill
\begin{minipage}[t]{0.48\textwidth}
\includegraphics[width=\textwidth]{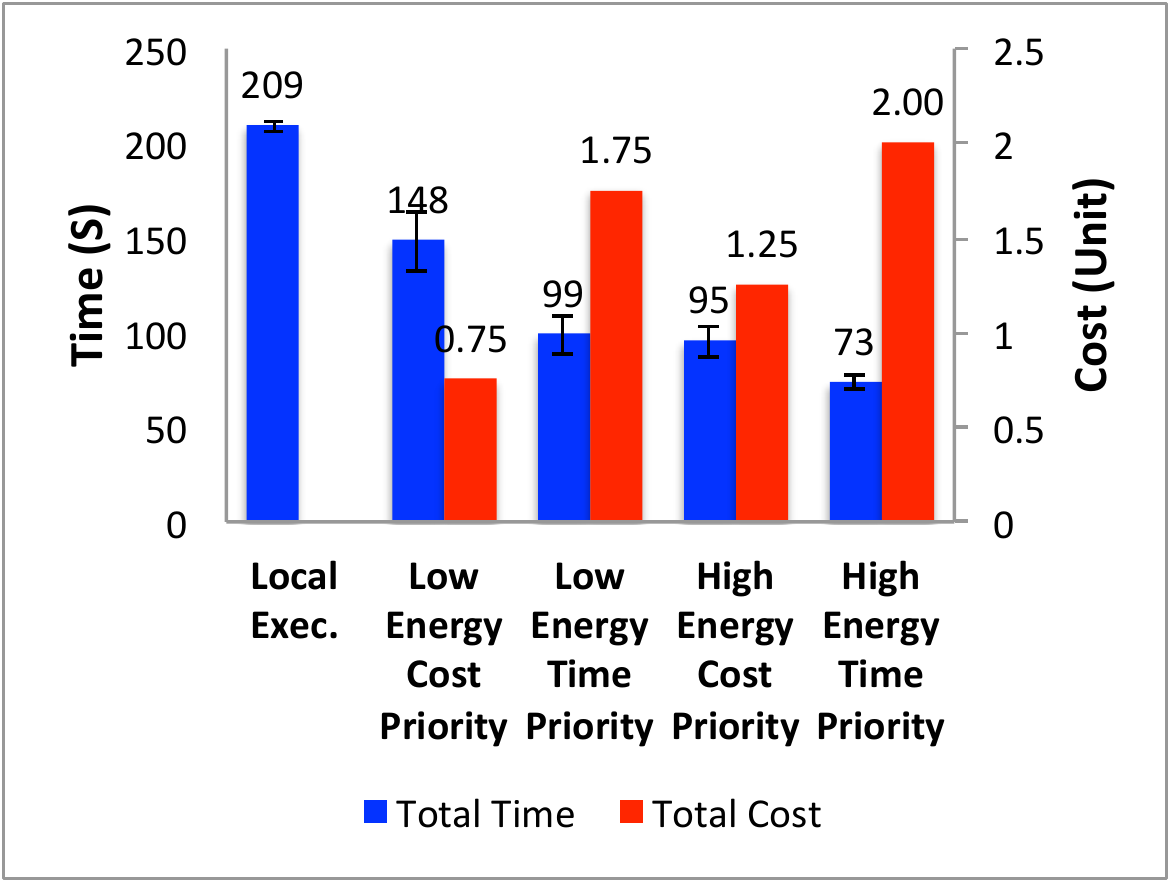} 
\caption{Impact of the collaborators' energy budget (3 collaborators \& 1 personal low end device).}
\label{fig:energy} 
\end{minipage} \hfill
\end{figure*}

We now evaluate the capability of Panorama to optimize for different objectives under different constraints specified by the participants. Consider the Bus or the Shopping Mall scenario with three mobile device collaborators and an additional mobile device (which is a tablet in the experiment) that belongs to the initiator. This corresponds to a situation where the initiator has two mobile devices, one of which is a low end device that has poor performance but is ``free'' in terms of cost and energy. The user may want to use the low end device to save time and energy or meet a cap on cost. We consider a situation where the initiator pays the collaborators, and the payment is proportional to the amount of work done.
%see Subsection ``Optimization Models''. 

%We envision here a scenario where the user would like to benefit from a nearby mobile node, say a tablet, that he or she owns. In the first experiment, we assigned a cost function for every device from the three devices including the initiator and assumed that the tablet is free of cost. The initiator was given a cost since we envision the user to want to minimize usage of the the mobile device to save energy or computing resources and shift the burden to the tablet node. One thing we would like to emphasize before describing the results is that the device used as the additional user node is a low end tablet that has slow performance. 

\begin{figure}
\centering
\includegraphics[scale=0.44]{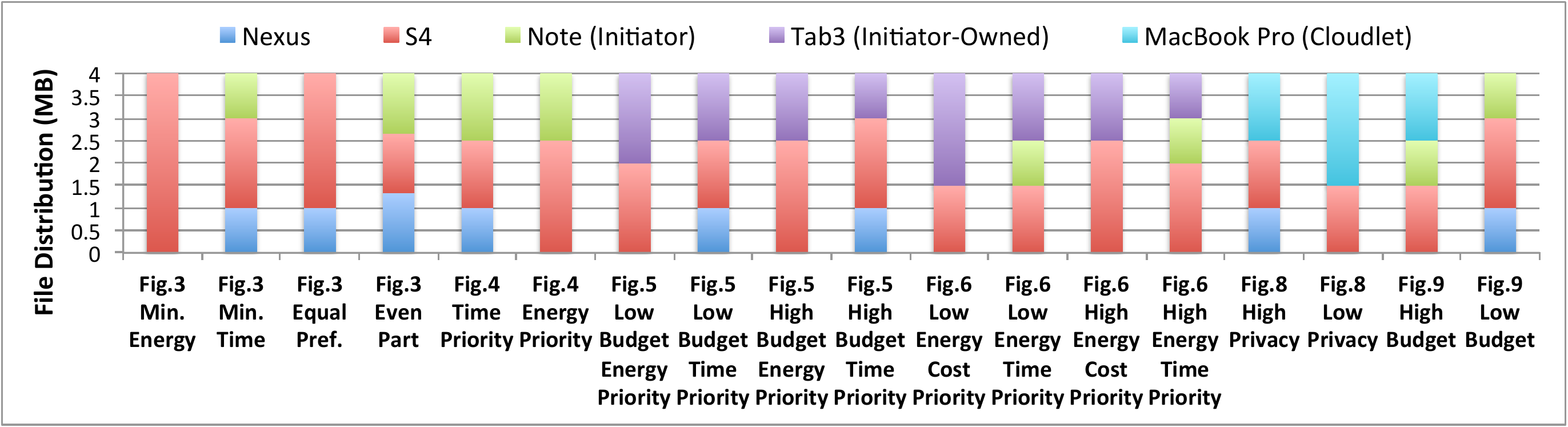}
\caption{File distribution for the 4MB task in experiments of Fig. 3, 4, 5, 6, 8 and 9.}
\label{fig:strategy-partition}
\end{figure}

We first consider a scenario in which the initiator has a budget on the total amount she is willing to pay. We experiment with two budget levels:
a low budget of 2 units of payment and a high budget of 
4 units of payment.\footnote{Notice that here the word ``payment'' is used in a general sense. It can be a monetary payment, or virtual payment such as credit for reputation.}
We consider two situations, one that aims 
%the initiator's device is low on the remaining
%battery level and so 
to minimize energy consumption, and
the other that aims  
%initiator's device has high level of remaining battery and so the 
to minimize execution time. 
The preference/priority on energy or time is represented by choosing
a small or large weight $\gamma$ in problem \eqref{eqn:obj2}-\eqref{eq:coneb1}.
Figure
\ref{fig:incetives} shows the results of this experiment, and the corresponding partitioning can be found in Figure \ref{fig:strategy-partition}.
%We optimized for energy and time while varying the user willingness to pay incentives. In the hard incentives scenario, the user assigns three incentives units to the collaboration. Whereas, in the relaxed incentives scenario, we doubled the paid incentives to six units. Units here can be substituted with any possible incentives mechanism such as micro payments or reputation points. Time and energy priorities where changed in the experiment by employing more weight for the objective with higher priority. 

We see that, with low budget and if energy is of high priority, Panorama sends most of the task to the free initiator-owned low end device, in order to save energy in other devices and meet the payment cap while incurring a long execution time.  When execution time is of high priority, Panorama sends the task more to the devices that are fast, which leads to $25\%$ reduction in time while costing much more energy.  On the other hand, with high budget, if time is of high priority, Panorama achieves a large reduction in time by shifting larger portion of job to fast but costly devices. When energy is of high priority, the energy consumption goes up compared to the low budget case. This %seems counter-intuitive, but 
is because the large reduction in time from faster computing allowed by higher budget compensates the increase in energy consumption. As expected, compared to local execution, collaboration reduces execution time and saves energy. %\lijun{In Figures 4 and 6, total energy means the energy in the initiating device or all the energy in collaborating devices other than the personal low-end device?} \khaled{all devices including initiator. Only energy on low end device is not included because this device is assumed ``free''}

%This in turn, caused huge saving in terms of energy compared to the other scenarios. Also, when hard incentives were given but time was more important for the user, Panorama utilized the given incentives to shift some file to the other costly users, which produced gain in terms of execution speed but caused energy consumption to increase. Now, we turn to the case where the initiator provided higher incentives for the collaboration. When time is set as a priority, we notice a huge gain in time due to shifting bigger portions to high speed but costly nodes. A result that we would like to highlight is that when given higher incentives and optimizing for energy, the energy consumption went up. This counter intuitive behavior happened because of two reasons. First, the saving in energy happens when shifting bigger portions to the file to the user owned node, which is already a free node. Second, although time is less priority than energy, the huge gain in time, 412 seconds compared to 752 seconds, compensated for the energy loss and caused Panorama to choose this distribution structure.

We now consider a scenario where the collaborators have a restriction on the amount of energy they are willing to expend for collaboration, and investigate the tradeoff between execution time and initiator's cost/payment under different energy budgets; see problem \eqref{eqn:obj4}-\eqref{eq:conwl1}. Figure \ref{fig:energy} shows the results of an experiment with a low, 100 Joules energy budget for each mobile device, and a high, 200 Joules energy budget for each device; and the corresponding partitioning of speech data can be found in Figure~\ref{fig:strategy-partition}.  %shows the corrsponding partitioning of speech data for each of the collaborators for these different cases. 
%The preference/priority on cost or time  is represented by choosing a small or large weight $\gamma$ in problem \eqref{eqn:obj4}-\eqref{eq:conwl1}, respectively. 
We see that, compared with low energy budget case, high energy budget leads to shorter execution time when comparing both the cost and time priority cases to their corresponding low energy cases. This is because higher energy budget allows for longer use of faster devices. Also, notice that, with low energy budget and if the cost is of high priority, Panorama sends larger portion of the task to the free initiator-owned low end device, resulting in large execution time and the lowest cost. %\lijun{In Figure 5, for local execution, there should be no cost. Did you set the payment $c_1$ to the initiating device to zero in all the experiments (without loss of generality, we can assume $i=1$ denotes the initiating device)? Same problem with Figure 6.} \khaled{I assumed that the initiator device also has a cost. I looked at it as if the initiator have a budget and can either spend it locally or with others depending on the objective} 
We also compare with the case of local execution. As expected, collaboration leads to shorter execution time while incurring cost as a result of utilizing other nodes.
%except for the case with low energy budget and cost priority that has a longer execution time. This is because a large portion of job is sent to the low end device, in order to meet the energy budget and reduce cost. 
%his result is intuitive since with limited energy, Panorama's ability in exploiting parallelism was limited by the willingness of the nodes to contribute energy. Another thing to note from the results that with limited energy and while prioritizing for cost, Panorama performed the same behavior of shifting the cost to free, but slow, user-owned node. In this situation, the paid incentive is very low, which is in harmony with the user overall objective of saving incentives.

\subsubsection{Presence of Cloudlets} 
\label{subsect:pc}

\begin{figure*}[!t] 
\begin{minipage}[t]{0.48\textwidth}
\includegraphics[width=\textwidth]{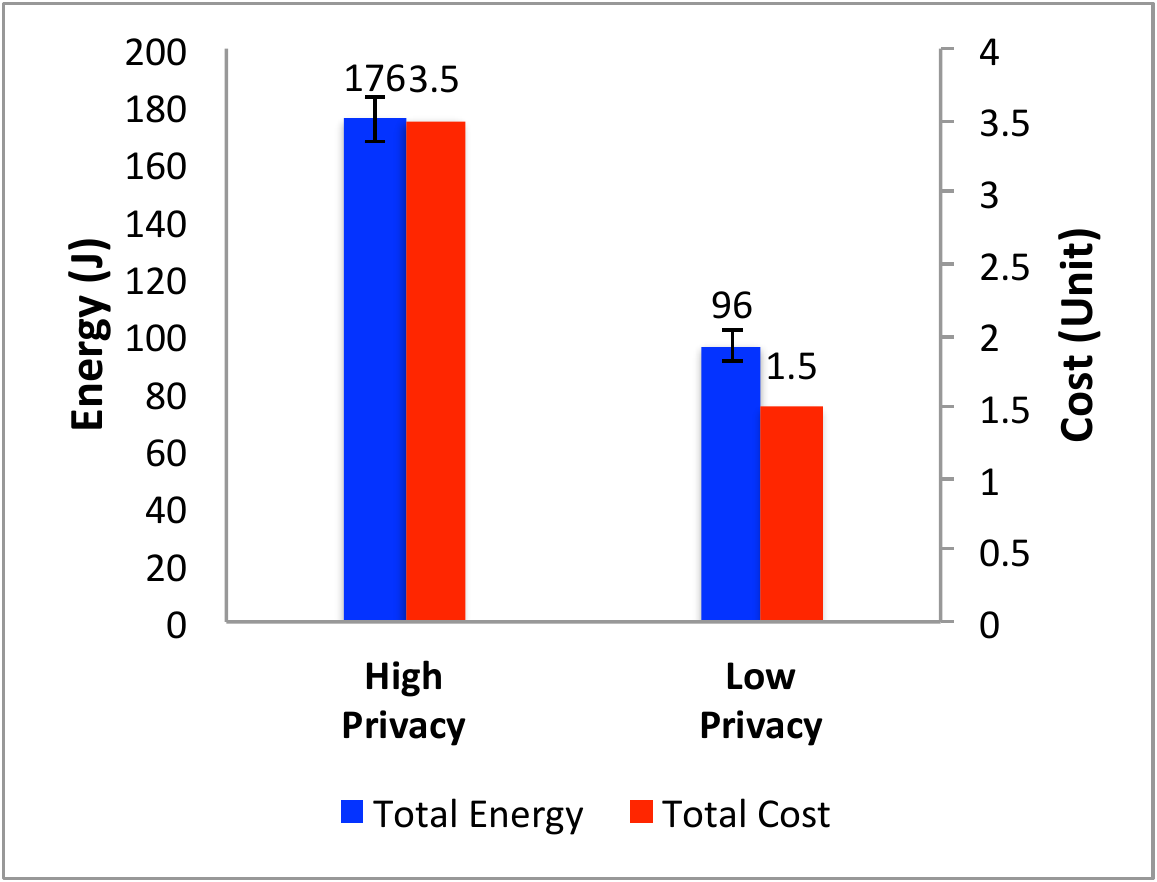} 
\caption{Impact of privacy requirement of the task (3 collaborators \& cloudlet).}
\label{fig:privacy} 
\end{minipage} \hfill
\begin{minipage}[t]{0.48\textwidth}
\includegraphics[width=\textwidth]{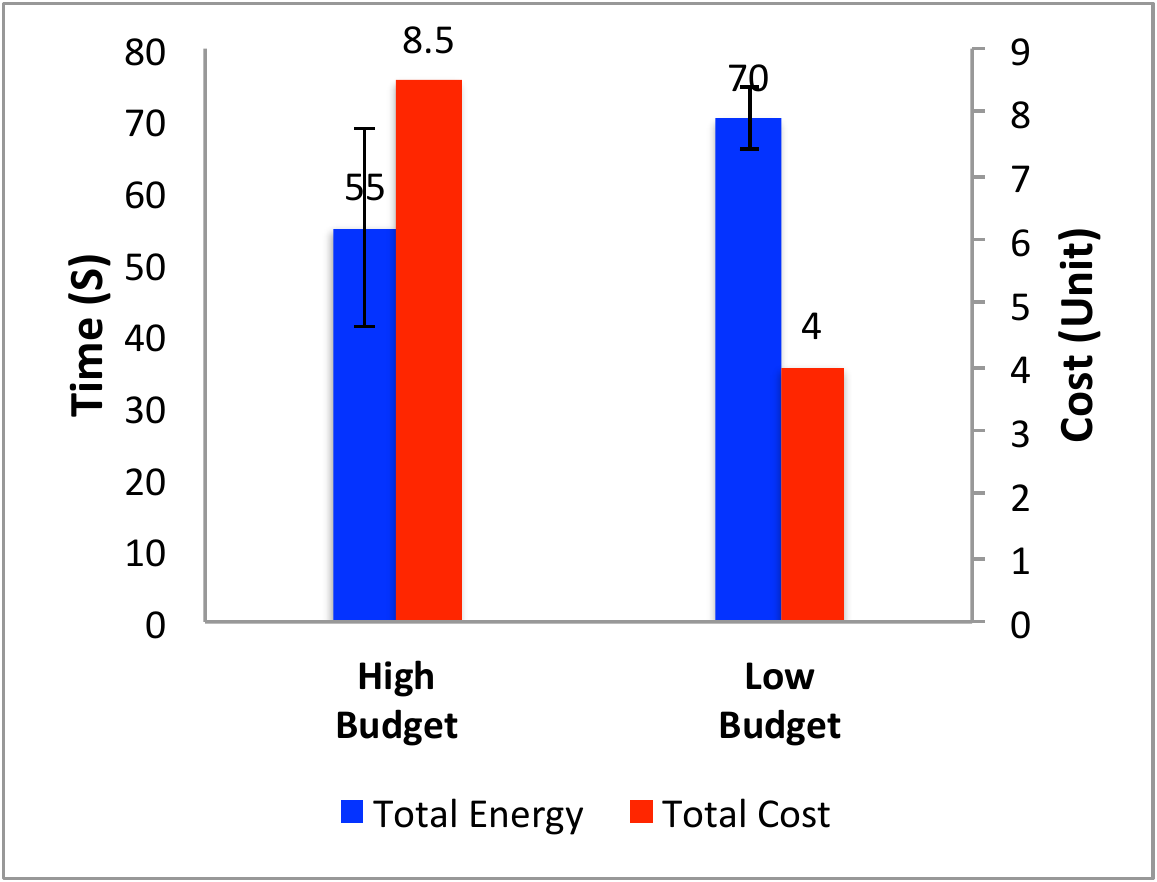} 
\caption{Impact of collaborators cost budget requirement of the task (3 collaborators \& cloudlet).}
\label{fig:budget} 
\end{minipage} \hfill
\end{figure*}

\comment{
We next consider the case where the speech data has some sensitive
information.  A major obstacle for systems involving task offloading like Panorama is
the risk of breaching the user privacy. Unlike time, energy, and incentives,
privacy is difficult to quantify. Devices other than the initiator's are
either trusted or not, and the data to process is either sensitive or not.
Here, we envision a scenario where a task consists of subtasks or data with
some parts being more sensitive than others. This can be images with faces
detected for image processing, or parts where users are likely to say
passwords or Personally Identifiable Information (PII) like the beginning
of a phone call for audio processing. With the presence of multiple
potential collaborators, Panorama can devise a partitioning plan that sends
sensitive subtasks and data to trusted nodes and non-sensitive ones to
less trusted nodes while optimizing for generic objectives such as energy
and time (See the discussion in the second last paragraph of Subsection ``Optimization Models''). 
}

We now consider scenarios where a cloudlet is available in addition to
some mobile devices for collaboration, as in the Work Place and Lunch Break
scenarios shown in Figure~\ref{fig:typical}. In a Work Place scenario,
a user has high trust in the available mobile devices as they belong to
her/his co-workers. In addition, in some Work Place scenarios, the user may
also trust the cloudlet, while in other cases, she/he may not trust it.
On the other hand, in the lunch break scenario, neither the cloudlet nor the
other mobile devices may be trusted.

In the first experiment reported here, we consider a Work Place scenario that
involves three mobile devices and an untrusted
cloudlet. An audio file is divided into sensitive and insensitive parts. 
We consider two cases here: a high privacy case with 2.5 MB out of the
total 4 MB file marked as sensitive, and a low privacy case with only 
1.5 MB marked as sensitive.
%To map sub tasks to resources according to sensitivity, Panorama performs planning twice. First, it plans insensitive parts to low trusted nodes. After that, it plans highly sensitive data to trusted nodes. Note that Panorama uses allocations of the first optimization as input to the second optimization. This ensures considering time and energy consumption of the first optimization when allocating sub tasks in the second phase. 
As shown in Figure~\ref{fig:privacy} (and Figure~\ref{fig:strategy-partition}), imposing higher privacy leads to higher energy consumption and higher cost. This is because only a small portion of the speech data is sent to the faster and energy-cost-free cloudlet. For lower privacy case, a much larger portion of the speech data is sent to the cloudlet, thus reducing energy consumption and execution
time. We conduct two additional experiments for the cases when the cloudlet
is trusted and when there is no sensitive data in the audio file. In both 
cases, Panorama offloaded almost the entire file to the 
cloudlet. This is because the cloudlet is significantly faster than the mobile
devices and does not contribute to energy overhead. We do not report the
results of these experiments here due to space limitation.

Next, we consider the Lunch Break scenario where both the cloudlet and the collaborating mobile devices are untrusted and there may be a cost associated with using them. In such a situation, the user has no choice other than executing sensitive parts of the task locally. Yet, Panorama can still optimize for the remaining non-sensitive portion of the task to devise an efficient plan, thereby, minimizing the burden on the initiator as much as possible. Figure~\ref{fig:budget} reports the results of an experiment where the user would like to process 4 MB of insensitive speech while minimizing the execution time (i.e., $\gamma=\infty$); and the corresponding partitioning of data can be found in 
Figure~\ref{fig:strategy-partition}. In contrast to the previous experiment, we gave the cloudlet here a higher cost of 4x compared to 1x for other nodes. %including the initiator node.  
In the case ``high-budget,''  the user allocates a budget of 10 units to the task, whereas, in the ``low-budget'' case only 5 units are allocated. We see from the figures that when the budget is high, Panorama was able to shift a big portion of the task to the cloudlet achieving better computation speedup compared to low budget scenario. However, the low budget scenario was not as slow as we expected when compared to the high budget scenario. The reason is that Panorama was able to exploit parallel execution (see Figure~\ref{fig:strategy-partition} for file distribution) with other collaborators within the allocated low budget without worrying about energy consumption since it is not considered in this scenario.

%For better comparison, we contrast these two scenarios with the local execution scenario. We see that even when the budget is low, Panorama achieved $50\%$ reduction in time compared to local execution by shifting parts of the task to the cloudlet and other mobile nodes and exploiting the parallel nature of the task. When the budget is high, the reduction in time increased to $56\%$. Clearly, Panorama's flexibility can bring significant reductions by utilizing available opportunities while still meeting the user's budget constraint. 
%In particular, we assume that the audio file is divided into \bf 1 MB of sensitive data and 3 MB of non-sensitive data.} In the case of Bus scenario, the user has low level of trust in the mobile devices of the collaborators as those collaborators are strangers. However, in the case of Shopping Mall scenario, the user has high level of trust in the mobile devices of the collaborators as those collaborators are family members. Figures~\ref{fig:strategy-privacy} and~\ref{fig:strategy-partition-privacy} show the energy-time tradeoff and partitioning provided by Panorama. {\bf Discuss details here}. {\bf These two experiments demonstrate that Panorama does a good job of partitioning to realize energy time tradeoff under current constraits.}
%\textcolor{blue}{We need to discuss details of these experiments}
%\input{infrastructure}
\subsection{Benefits of Collaboration for Sequential Tasks}
\begin{figure*}[!t]
%\begin{minipage}[t]{0.32\textwidth}
%\includegraphics[width=\textwidth]{Figures/WhenToOffload.pdf} 
%\caption{Benefits from various collaboration opportunities}
%\label{fig:when} 
%\end{minipage} \hfill 
\begin{minipage}[t]{0.48\textwidth}
\includegraphics[width=\textwidth]{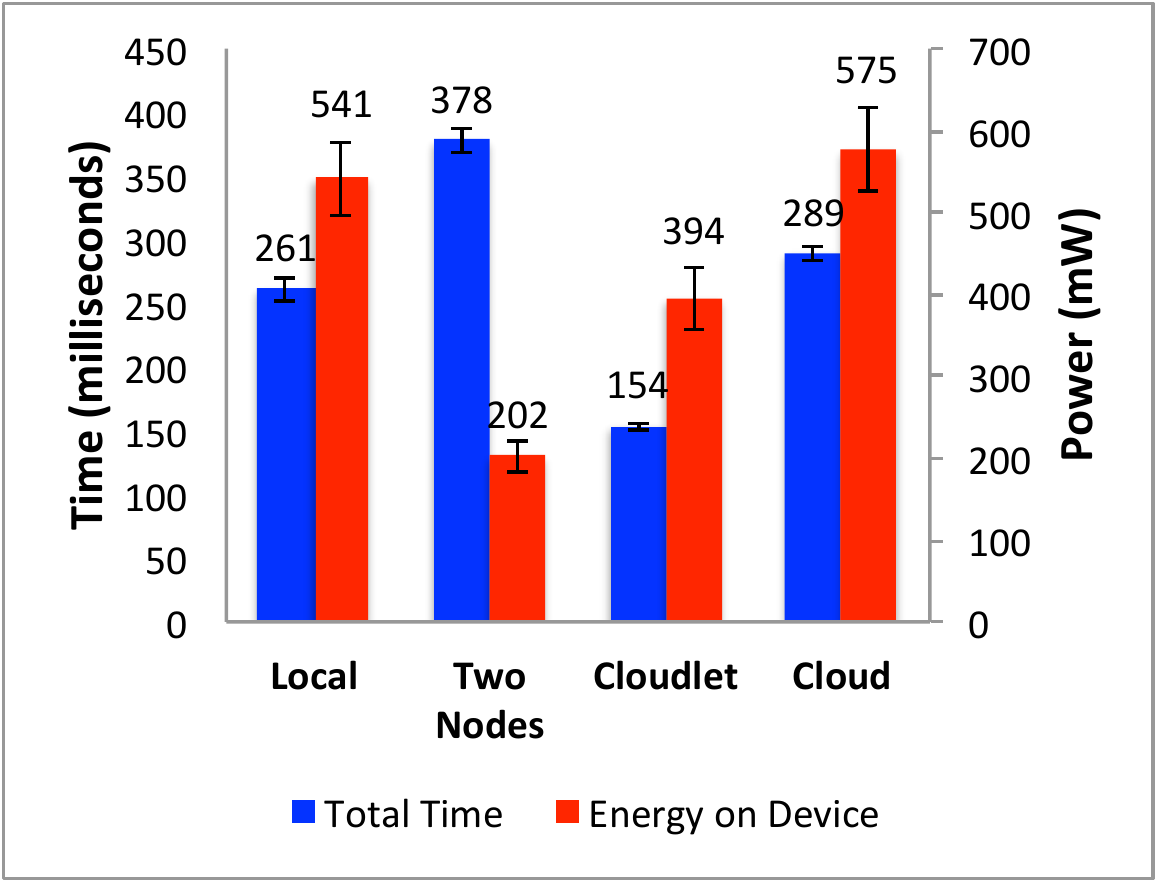} 
\caption{Benefits of collaboration for sound ambiance monitoring.}
\label{fig:sequential} 
\end{minipage} \hfill
\begin{minipage}[t]{0.48\textwidth}
\includegraphics[width=\textwidth]{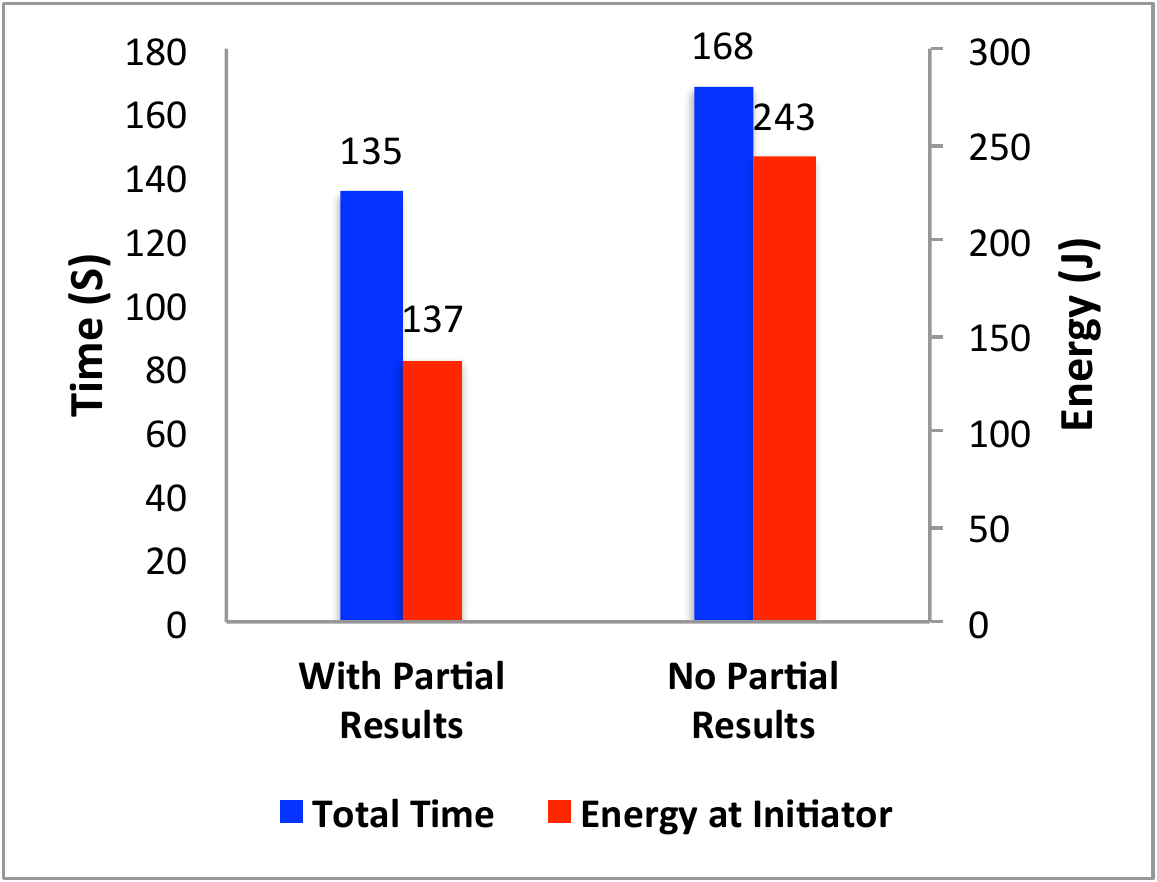} 
\caption{Impact of leaving node on Panorama's performance.}
\label{fig:moving} 
\end{minipage} \hfill
\end{figure*}

To demonstrate the utility of Panorama in handling sequential task structures, we have implemented the sound ambiance monitoring application proposed in \cite{lu2009soundsense}, and employed Panorama to enable collaboration. The results of the collaboration experiments are then compared to the local execution on a single device with Panorama turned off. %Figure \ref{fig:sequential} shows the completion time and power consumption. 
For collaboration, we conducted three experiments. In the first experiment, the initiator collaborates with two other mobile devices, and in the second and third experiments, the initiator collaborates with a cloudlet sitting on the same network and a cloud server accessed through a Wi-Fi Internet connection. Recall from Section \ref{sec:testbed} that the sound ambiance monitoring task can be viewed as a pipeline consisting of three subtasks. Those can be split among collaborators. 
%In the first stage, it captures a three-second audio from the microphone of a mobile device. In the second stage, it calculates the Fast Fourier Transform (FFT) from the time-domain data of the captured audio. It then calculates features based on the FFT data and uses these features to classify the ambiance sound in the recorded three seconds as music or speech. Finally, in the third stage, it calculates Mel-frequency cepstral coefficients (MFCCs) based on the FFT data taken as input from the second subtask. It then uses the coefficients to classify speech windows according to the speaker gender.
% where we report the average and the standard error of five readings observed during the continuous processing of audio frames to perform the online sound ambiance classification.
From Figure \ref{fig:sequential}, we see that in case of collaboration with two other mobile devices, there
is a reduction in power consumption at the initiator's device from 541 mW to 202 mW, a more than $50\%$ energy reduction by delegating the calculation to the other collaborator. However, this comes at the cost of an increased completion time from 261 milliseconds to 378 milliseconds. Completion time here is the time between when the audio recording is completed and when the gender classification result has arrived at the initiator from the collaborator (or calculated locally in case of local execution). The increase in completion time is due to the time needed to set up the collaboration task and transfer the data and results between the collaborators.

Interestingly, the second experiment involving cloudlet does not show an increase in completion time. Instead, a time saving of $40\%$ is achieved in addition to the energy saving of $27\%$. Here, the overhead introduced by Panorama
%to divide and ship the subtasks 
is offset by the significant gain in execution time when delegating the compute-intensive parts to the cloudlet. We also ran the same experiment to engage in a collaboration with an Amazon EC2 sever over a Wi-Fi Internet connection. The achieved result is worse both in terms of energy and time when compared to the cloudlet case. However, when we compare this result to a nearby node collaboration, we see that cloud can be a better alternative, depending on the intensity of the task, in terms of time while the opposite was true for energy.
% in this experiment.

\subsection{Handling Mobility}
%Panorama uses short-range communication channels to connect mobile devices during collaboration. This causes the collaboration network to be susceptible to disconnection due to node mobility. 
There are two main challenges when it comes to handling node mobility: how to detect that a node is moving away, and how to ensure smooth migration of unfinished sub-task to other devices when a node leaves. For detection, we use a method proposed in \cite{miluzzo2012vision} where we sense the accelerometer during collaboration to detect the starting of a physical activity as an indicator for a collaborator eventually leaving the scene. Once Panorama detects such activity, it sends a message to the initiator to handle mobility. We use accelerometer due to its relatively cheap energy cost and the fact that it can detect mobility promptly.
%Other approaches that we can use to detect a node moving away are heartbeat messages and considering late nodes as stragglers and performing speculative executions analogous to MapReduce.
%To take care of unfinished subtask, in the current implementation of Panorama we simply let the initiator execute it. We may also optimally partition the unfinished subtask of a leaving node to the remaining collaborators. However, we believe this will lead to significant increase in total execution time. 
Handling of interrupted collaborations depends heavily on the nature of the computation. In some situations a partial result can be migrated back to the initiator, whereas in others the whole computation need to be reprocessed. We performed two experiments with the aforementioned cases and report the results in Figure \ref{fig:moving} to reflect the impact of each on performance. The experiment considers 
a scenario where a node delegates a 4 MB task equally to two other nodes and one node moves away from the initiator. In the first case, we deliberately divided the received file and let the moving node finish the first 1 MB before moving away. Upon moving away, the node sends the partial result back to the initiator, which processes the remaining 1 MB locally. In the second case, we let the moving node report its movement without sending any partial results, so the initiator will process the whole 2 MB. Figure \ref{fig:moving}  reports the total time for completing the 4 MB task and the consumed energy at the initiator. As expected, the first scenario of partial result migration is better in terms of both energy and time when compared to the second scenario, since the initiator only needed to process half of the load assigned to the moving node.
Notice that in our current implementation, when a collaborating node is leaving, the initiator picks up the unfinished work. We can also re-distribute the unfinished work among the remaining devices, which we plan to explore in future. 
%Figure \ref{fig:moving} shows the results of an experiment that compares the situations with and without a node leaving.  As expected, node mobility leads to an increase in both energy and time. 
%In fact, it is possible to incorporate node mobility as an input in Panorama's 
%Multi-Objective Optimizer. The key idea is that a node monitors its
%current mobility and reports it as low, medium or high to the task initiator.
%The task initiator then incorporates this in its optimzation process. We
%plan to experiment with this idea in future.
%However, Panorama %We performed an optimization involving three mobile nodes and repeated the same experiment while moving one node away. We note that a penalty of increased execution occurred since the initiator was busy executing their own assigned sub task. Hence, the re-execution of the lost portion need to wait for that sub-task to finish. As for energy, consumption was not much higher since we didn't involve the wasted cost at the moving node.

\section{Related Work}
\label{sec:related}
%Panorama fits under the umbrella of solutions that utilize external resources to augment the capabilities of a mobile device. Specifically, we employ this computing paradigm to tackle the challenging requirement of continuous context monitoring in context-aware applications. 
%Several different techniques at the mobile device level have been proposed to enable efficient continuous sensing; see,  e.g., \cite{wang2009framework,kang2008seemon,ju2012symphoney}. Although our goal is similar towards supporting continuous context monitoring task, we target collaboration among mobile devices and
%offloading to cloud/cloudlet. We believe that although the idea of collaboration between co-located mobile devices has been discussed in many existing works, there is a lack of a concrete implementation showing how this computing paradigm can bring benefit to a specific application, as well as a lack of a mathematical modeling framework for guiding optimal collaboration planning. This paper fills this gap.
% by designing and prototyping Panorama and demonstrating its effectiveness and efficacy by applying it to two representative context-aware applications. %and fills this gap by applying co-located mobile devices collaboration in the context of continuous user context monitoring and 

Our work is closely related to \cite{lee2012comon,miluzzo2010darwin}. However, \cite{lee2012comon} focuses on collaborative context monitoring between co-located mobile devices only, while Panorama leverages  more opportunities by involving not only co-located devices but also cloudlets/cloud and performing an optimization to devise an optimal collaboration plan for the task. The main goal of \cite{miluzzo2010darwin} %looked at utilizing co-located mobile devices to serve continuous context monitoring. However, the main goal of that work 
is to enhance the reliability of the application, while the goal of Panorama is to automate and optimize collaboration for continuous context monitoring.
% to pave the way for wide adoption of this kind of applications. 
%Also, collaboration between co-located devices has been used in \cite{keller2012microcast} to provide efficient video streaming. by utilizing communication resources of the devices involved in the collaboration.

The work in \cite{shi2012serendipity} studies generic computation offloading between co-located mobile devices, and presents three algorithms to serve three different possible applications' structures while taking into consideration connectivity in distributing jobs. The implementation in \cite{shi2012serendipity}  is limited to a prototype that performs offloading between two devices only. Panorama's design involves more opportunities by including cloudlets/cloud in addition to co-located mobile devices. We also provide an extensive Android implementation and evaluate it on multiple mobile nodes and a cloudlet/cloud.
The recent work in \cite{de2015group}
%work that proposed utilizing collaborations between mobile devices to calculate context is presented in \cite{de2015group}. This work
focuses on building a conceptual model to facilitate context sharing between groups of mobile devices. Such model can be leveraged by Panorama to increase the chances of meeting peers and building more beneficial collaborations.

There are several vision papers that advocate the concept of collaboration among co-located mobile devices \cite{miluzzo2012vision,conti2010opportunities,zhang2013toward} and we have used some of their ideas to motivate our work. Also, a rich body of literature exists for augmenting smartphones with resources from cloud and cloudlets; see, e.g., \cite{cuervo2010maui,satyanarayanancloudlets,ra2011odessa,chun2009augmented}. The ideas in these works have helped in guiding our design.

\section{Conclusion}
\label{sec:conclusion}

%Offloading techniques have been proposed over the last several years now to
%address the problem of power and compute constraints of small mobile devices.
%These offloading techniques have their own overheads in terms of discovering
%appropriate offloading opportunities, transfering code and data, and getting 
%the results back. In addition, multiple offloading opportunities in the
%form of other mobile devices, cloudlets and cloud may be available,
%and the collaborators themselves may have a variety
%of constraints in terms their energy and processing quota, security and privacy
%issues, access cost, and mobility. In light of this, a key question is when
%should a device offload its context computing task and how.  
%This paper presents Panorma that addresses this key
%question.

Panorama is a middleware framework that 
addresses a key question in offloadling computations to nearby mobile devices
and cloudlets/cloud: when
should a device offload its context computing task and how? Panorama utilizes all available collaboration opportunities from co-located mobile
devices and cloudlets/cloud, and devises a collaboration plan to optimize
for and trade off different objectives such as minimizing execution time or 
minimizing energy consumption. The optimization algorithm
considers limits set by participants such as contributed energy, paid
incentives, and privacy exposure. 
%The paper provides an extensive evaluation of Panorama using two different
%context monitoring applications over four mobile devices, a cloudlet and a
%cloud server under several different combinations of device constraints and
%multiple objectives. 
Evaluation results show that Panorama is rather
practical, is able to cope
up with varying device constraints, and devises collaboration plans
within those constraints to optimally trade off multiple objectives. 

There are a number of future directions we plan to pursue. First, we plan to
incorporate Bluetooth Low Energy in our opportunity discovery protocol. While
none of the Android devices we test run in peripheral mode at present, we
expect that Bluetooth-enabled smartphones will increasingly support
Bluetooth LE. 
%This will likely reduce power consumption significantly during
%discovery process. 
Second, we plan to expand on handling node mobility.
At present, Panorama provides basic support for ensuring that the
context computation task is completed despite some of the devices moving away.
We plan to explore smart ways to efficiently cope with various mobility patterns. 
%We plan to explore energy-efficient, smart ways to detect device mobility and execute the remaining subtask that is assigned to the leaving device.
Third, a limitation in Panorama is to rely on collaborators to come up with privacy and efficiency requirements. An interesting research direction we plan to pursue is to automate the process of generating these requirements to enhance the practicality of Panorama. Finally, we plan to conduct user studies to evaluate Panorama in the real-world setting.

\bibliographystyle{plain}
\small
\bibliography{sigproc}

\end{document}